# A Generalized Landauer Approach for Electron Transport in a Single Molecule

Augusto C. L. Moreira and Celso P. de Melo*
Departamento de Física, Universidade Federal de Pernambuco
50.670-901 Recife, PE, Brazil

**ABSTRACT**

We present a quaternion-inspired formalism specifically developed to evaluate the electric current that traverses a single molecule subjected to an externally applied voltage. The molecule of interest is covalently connected to two small metallic clusters, forming an "extended molecule" complex. The quaternion approach allows for an integrated treatment of the charge transport in single molecules where both ballistic and co-tunneling (coherent) mechanisms are taken on equal footing, although only in the latter case the presence of eventual transient charged states of the system needs to be considered. We use a Dyson series to obtain a generalized Fermi golden rule, from which we derive an expression for the net current the two electrodes: in doing this, we take into account all possible transitions between electronic states localized at the electrodes and levels in the "extended molecule" complex. In fact, one can apply the method to the entire range of coupling regimes, not only in the weak or strong cases, but also in intermediate situations, where ballistic and co-tunneling processes compete with each other. We also discuss initial results of the application of this formalism to the description of the electronic transport in two small organic molecules representative of two different limit situations. In the first case, a conjugated molecule (where spatially delocalized molecular orbitals favor ballistic contributions) is considered, and in the second the current traverses a saturated hydrocarbon (whose structure should contain more localized molecular orbitals). In both cases, we fully describe the field-induced self-adjustment of the electronic levels of the extended molecule complex at an *ab initio* quantum chemical level, using density functional theory.

# I. INTRODUCTION

Since Aviram and Ratner first proposed the concept of a molecular rectifier,[1] a crescent amount of theoretical effort has been dedicated to the understanding of electronic transport processes occurring at the molecular scale.[2] However, after so many years of hard and varied effort, the calculation of the current that flows through a single molecule connected to metal electrodes under an externally applied voltage remains one of the most difficult theoretical problems to be properly considered.

Molecular electronics is expected to present many advantages when compared to present day silicon based technology, such as the possibility of preparing devices of smaller size, higher speed, lower cost and novel functionalities.[2] In principle, a single molecule can play different active roles in a device, but a few general conclusions can be safely inferred, such as: *i)* long molecules have lower conductance than short ones, *ii)* conjugated systems have higher conductance that non-conjugated molecules of similar size, and *iii)* molecules with structural asymmetry exhibit asymmetric current-voltage characteristics. The fact that one can relate the intrinsic molecular structure of the system to the measured current and conductance suggests that it is feasible the theoretical modeling of these two latter quantities.

# II. CONDUCTANCE THROUGH SINGLE–MOLECULES

## A. Transmission function

A first theoretical insight into the question of the mechanisms involved in the molecular transport appeared with Landauer's introduction of the concept of 'conductance from transmission'.[3] The electric current *I* of a very small conductor is associated to the coherent probability of transmission *T(E)*

$$I_{coh} = \frac{e}{\pi\hbar} \int_{-\infty}^{\infty} dE\, T(E)\left(f_L(E) - f_R(E)\right) \quad , \tag{1}$$

through the internal transversal modes, which are supposed to be limited in number when compared to the infinite range of those possible in the external electrodes and in the Fermi distribution functions ($f_{L,R}(E)$) at the two (left/right) semi-infinite electrodes.[4]

Different suggestions on how to calculate *T(E)* have been introduced, the most successful of them based on scattering theory. For instance, by use of the Lippmann-Schwinger (LS) equation [4] $|\psi\rangle = |\phi\rangle + \hat{G}^+\hat{V}|\phi\rangle$, one can construct the exact scattering states $|\psi\rangle$ with basis in the knowledge of the original unperturbed state $|\phi\rangle$ and the potential *V* that couples the device region to the contacts, through the retarded Green´s function $\hat{G}^+$. In this case, the transmission *T(E)* is related[5] to the scattering matrix *S* that appears when we take the square of the modulus of $|\psi\rangle$. Another possibility is to adopt the so-called Non Equilibrium Green´s Function (NEGF) formalism.[4,6]

In the former type of approach, the molecular coupling to the two terminal contacts (which are labeled as *L* and *R*, respectively) is described by an effective device Hamiltonian $\hat{H}^{eff} \rightarrow \hat{H} + \hat{\Sigma}_L + \hat{\Sigma}_R$ that incorporates each electrode by taking into account its corresponding self-energy matrix $\hat{\Sigma}_i$ (i = L, R). Once this effective Hamiltonian is known, one can calculate the overall current through the system. In the non-interacting limit case of the NEGF, the transmission can be expressed as[4]

$$T(E) = Tr\left\{\hat{\Gamma}_L(E)\hat{G}^+(E)\hat{\Gamma}_R(E)\hat{G}^-(E)\right\}. \tag{2}$$

In Eq. (2), $\hat{G}^+$ and $\hat{G}^-$ are the retarded and the advanced Green's function of the coupled system, respectively, while $\hat{\Gamma}_i$, the broadening contribution of the molecular levels, is associated to the anti-hermitian part of $\hat{\Sigma}_i$.

Please note that the presence of an explicitly "transmission function" through the molecular system remains valid as long as the electron transport is of a "ballistic" nature, with no evidence of scattering at the molecular internal structure. In fact, this latter phenomenon is typically associated to the presence of spatially localized molecular states.[3] On the contrary, ballistic transport of electrons must necessarily involve a "hot electron" scattered by an unoccupied molecular orbital that should be spatially delocalized throughout the entire system.[7]

### B. Strong, weak and intermediate coupling regimes

The ballistic transport occurs whenever the interaction between the molecule and the metallic leads is strong. In the strong coupling limit, sometimes referred to as the self-consistent field regime[6] (SCF), the molecular levels are broadened by their interaction with the contacts. The corresponding energy (of the order of $k_B T$) is comparable to the single-electron charging energy $U_0$, i.e., the energy separation between states of different spin orientation. In fact, the value of the charging energy itself is related to the extent of the spatial localization of the wave function: levels with well-delocalized wave functions have a very small $U_0$.[6, 8] Under these conditions, even small external bias may cause a detectable current to traverse the device.

It is worthwhile to note that one can also use the above-mentioned methods to describe non-coherent phenomena (such as electron-phonon coupling) by adding an interaction potential and then evaluating the corresponding coupling constants. As shown by Datta,[3] while in the case of NEGF method this procedure would be equivalent to attach an extra "contact" to the device with a self-energy $\hat{\Sigma}_S$ and an additional coupling $\hat{\Gamma}_S$ representing the involved states, in the LS equation one makes the replacement $\hat{V} \rightarrow \hat{V} + \hat{U}$. In both cases, the complete wave function must now take into account both individual sub-spaces associated to phonon and electrons.

In the usual treatment of molecular electronic devices,[9-11] even after inclusion of non-coherent contributions the transport through the system remains confined to a single electronic potential surface; as a consequence, the active component (the single molecule, in our case) must remain unaltered in its original electronic state (be it its neutral form or a specific anion or cation). However, this picture is not valid if the coupling between the molecule and the metallic leads is sufficiently small; in this limit, the charge transport is sequential and we must allow for the presence of transient charged states of the molecular species considered.[6] One now faces the regime of weak coupling, where charging effects such as Coulomb blockade,[12] for example, may occur. In this case, either the multi electron master equation (MEME) approach must be employed or one could resort to NEGF methods out of the non-interactive limit, adding Coulomb electronic repulsion energy term in the Hamiltonian and employing the full expression for the electric current (see, for instance, Ref. [4], p. 242). In this weakly coupled regime (sometimes referred to as the Coulomb blockade (CB) limit), the charging energy $U_0$ is much bigger than both $k_B T$ and the coupling between the contacts and the molecular levels. Now, due to the high strength of $U_0$, very little current would flow when a small external bias is applied to the device. Once again, the charging energy is related to the extent of the wave function spatial localization: in the Coulomb blockade regime, one should expect to encounter more wave functions of a highly localized nature.[6, 8, 13] Actually, in realistic nanodevices as is the case of a single molecule attached to metallic electrodes, delocalized and well-localized wave functions will coexist[13] and, hence, both Coulomb blockade and ballistic processes may contribute to the observed current. In this case, however, one should note that, although NEGF methods can be used to describe the charge flow in all (i.e., weak, strong and intermediate) coupling cases, they do not take into account a self-consistent charge distribution during the tunneling; consequently, the Green´s function of the complete problem is generally unknown.[14] Hence, when interactions beyond mean field approximation are included, it is not a simple task to compute the current in actual systems within the NEGF formalism. Therefore, one usually employs the orthodox MEME picture only when broadening effects are absent.

It is especially important to note that the issue of deciding *a priori* on whether a given molecular system will exhibit a strong rather than a weak coupling to a pair of existing metallic electrodes is far from settled. A criterion usually adopted[6] is to assume that a strong coupling will be developed through the molecular system if the corresponding molecular orbitals are sufficiently delocalized to present no vanishing electronic densities at the opposite sides of the extended molecule.[15] Also note that this assumption precludes the existence of strong coupling – and therefore of the occurrence of ballistic transport – in saturated molecules, where spatial localization of the electronic density at each extremity usually occurs.[1, 16]

However, a logical quagmire rapidly develops if one stretches the argument further on to consider larger and larger extended molecular entities. While, on one hand, a progressive increase in the number of metallic atoms in the terminal clusters of the EM should improve the quality of the calculated results,[16] at the other, the larger the size of the EM, less likely it becomes to find states that are truly delocalized throughout the extended molecule region. To compound to these difficulties, if the theoretical treatment of the problem allows for a new *ab initio* calculation of the molecular system every time that the external electric field is adjusted to a new value, the spatial localization of the frontier molecular orbitals can change accordingly. Consequently, the very nature of the transport regime will vary along the actual calculation of the current profile.

So, a more complete formalism for describing quantum transport through a single molecule connected to two terminal electrodes has to account for the occurrence of not only the two limiting regimes previously described in a separated manner; rather, it must consider the possibility that both regimes exist at once and treat them on equal *a priori* footing. One might also expect that the case of an intermediate coupling regime[6] would automatically be included in these more elaborated theoretical approaches. Note that, according to the intensity of the externally applied electric field, in the intermediate regime either ballistic processes or alternative charge transfer pathways involving transient charged molecular states (some of them corresponding to non-coherent tunneling events at the two electrode-molecule junctions[17]) may dominate the overall charge transport.

### III. THE QUATERNIONIC FORMALISM

#### A. Molecular charge states as independent quaternionic sub-spaces

Suppose that an electric field of varying intensity exists between two semi-infinite leads connected to opposite ends of the extended molecule. Then, once the three possible pathways are independently considered for each fixed value of the applied potential, both ballistic and co-tunneling terms will arise naturally in the resulting expression for the overall electric current. To be able to do this in a simple and elegant manner, we will recur to the use of a *quaternion*-inspired formalism. Quaternions is a theoretical concept originally introduced by W. R. Hamilton in the XIX Century[18] that, in modern times, has been adapted by Adler[19] to quantum mechanical problems.

A quaternion is defined as a hyper complex number $q = a + ib + jc + kd$, where $a$, $b$, $c$, and $d$ are real numbers and $i$, $j$ and $k$ are imaginary units that observe the cyclic properties $i^2 = j^2 = k^2 = ijk = -1$ and $ij = -ji = k, jk = -kj, ki = -ik = j$. Note that by taking advantage of three non-commuting distinct imaginary units, one can write three time-dependent Schrödinger equations, each one corresponding to a different unperturbed hermitian Hamiltonian, as

$$i\hbar\partial_t |\phi_{n_1}(t)\rangle = \hat{H}_{01} |\phi_{n_1}(t)\rangle \quad \rightarrow \quad \hat{H}_{01} |\phi_{n_1}(t)\rangle = E_{n_1} |\phi_{n_1}(t)\rangle$$
$$j\hbar\partial_t |\phi_{n_2}(t)\rangle = \hat{H}_{02} |\phi_{n_2}(t)\rangle \quad \rightarrow \quad \hat{H}_{02} |\phi_{n_2}(t)\rangle = E_{n_2} |\phi_{n_2}(t)\rangle \quad (3)$$
$$k\hbar\partial_t |\phi_{n_3}(t)\rangle = \hat{H}_{03} |\phi_{n_3}(t)\rangle \quad \rightarrow \quad \hat{H}_{03} |\phi_{n_3}(t)\rangle = E_{n_3} |\phi_{n_3}(t)\rangle \quad .$$

It is easy to show that these three Hamiltonian operators do not commute and therefore they do not possess simultaneous time-dependent eigenstates. In this manner, each one of the three Hamiltonians must exhibit its own set of eigenstates. In the end, each eigenstate can be made orthogonal to all the others. We can take advantage of this fact, and write the time evolution of these Hamiltonians in terms of a single matrix expression, in the form

$$\begin{pmatrix} i & 0 & 0 \\ 0 & j & 0 \\ 0 & 0 & k \end{pmatrix} \hbar\partial_t \begin{pmatrix} \gamma_1 |\phi_{n_1}(t)\rangle \\ \gamma_2 |\phi_{n_2}(t)\rangle \\ \gamma_3 |\phi_{n_3}(t)\rangle \end{pmatrix} = \begin{pmatrix} E_{0,n_1} & 0 & 0 \\ 0 & E_{0,n_2} & 0 \\ 0 & 0 & E_{0,n_3} \end{pmatrix} \begin{pmatrix} \gamma_1 |\phi_{n_1}(t)\rangle \\ \gamma_2 |\phi_{n_2}(t)\rangle \\ \gamma_3 |\phi_{n_3}(t)\rangle \end{pmatrix}$$
$$= \begin{pmatrix} \hat{H}_{01} & 0 & 0 \\ 0 & \hat{H}_{02} & 0 \\ 0 & 0 & \hat{H}_{03} \end{pmatrix} \begin{pmatrix} \gamma_1 |\phi_{n_1}(t)\rangle \\ \gamma_2 |\phi_{n_2}(t)\rangle \\ \gamma_3 |\phi_{n_3}(t)\rangle \end{pmatrix} , \quad (4)$$

where the $\gamma_x$ (x = 1, 2 and 3) normalization factors must necessarily obey the condition $\gamma_1^2 + \gamma_2^2 + \gamma_3^2 = 1$. We define the (unperturbed) quaternionic wave function and matrices, respectively, by

$$|\mathbb{X}_{0,n}(t)\rangle \equiv \begin{pmatrix} \gamma_1 |\phi_{n_1}(t)\rangle \\ \gamma_2 |\phi_{n_2}(t)\rangle \\ \gamma_3 |\phi_{n_3}(t)\rangle \end{pmatrix} \quad , \tag{5.1}$$

$$\mathfrak{q} \equiv \begin{pmatrix} i & 0 & 0 \\ 0 & j & 0 \\ 0 & 0 & k \end{pmatrix} \quad , \tag{5.2}$$

$$\mathbb{H}_0 \equiv \begin{pmatrix} \hat{H}_{01} & 0 & 0 \\ 0 & \hat{H}_{02} & 0 \\ 0 & 0 & \hat{H}_{03} \end{pmatrix} \quad , \tag{5.3}$$

$$\text{and } \mathbb{E}_{0,n} \equiv \begin{pmatrix} E_{0,n_1} & 0 & 0 \\ 0 & E_{0,n_2} & 0 \\ 0 & 0 & E_{0,n_3} \end{pmatrix} \quad , \tag{5.4}$$

so that, in a more compact form, we have

$$\mathfrak{q}\hbar\partial_t |\mathbb{X}_{0,n}(t)\rangle = \mathbb{H}_0 |\mathbb{X}_{0,n}(t)\rangle \quad \rightarrow \quad \mathbb{H}_0 |\mathbb{X}_{0,n}(t)\rangle = \mathbb{E}_{0,n} |\mathbb{X}_{0,n}(t)\rangle \quad . \tag{6}$$

We then have three orthogonal Hilbert spaces, each comprising one of the mutually non-commutative Hamiltonians. We will now discuss what happens when a time independent perturbation is present. Consider that a generic time independent perturbation like

$$\mathbb{V} = \begin{pmatrix} \hat{V}_{11} & \hat{V}_{12} & \hat{V}_{13} \\ \hat{V}_{21} & \hat{V}_{22} & \hat{V}_{23} \\ \hat{V}_{31} & \hat{V}_{32} & \hat{V}_{33} \end{pmatrix} \tag{6}$$

is introduced in the system, so that one can write $\mathbb{H} = \mathbb{H}_0 + \mathbb{V}$. As usual, we assume that each of its elements is much smaller than the characteristic energies of the unperturbed system. We can write a LS equation for this perturbed quaternionic system and so define a corresponding $\mathbb{S}$ matrix (and, consequently, a $\mathbb{T}$ matrix). To see this, assume that the eigenvalues of $\mathbb{H}_0$ are known, such that $\mathbb{H}_0|\mathbb{X}_{0,n}\rangle = \mathbb{E}_{0,n}|\mathbb{X}_{0,n}\rangle$, and consider the Hamiltonian $\mathbb{H} = \mathbb{H}_0 + \mathbb{V}$. In the usual treatment, one associates $\mathbb{H}_0$ to a free particle, while $\mathbb{V}$ is a scattering potential. The corresponding LS equation can be obtained if one admits that each unperturbed solution $|\mathbb{X}_{0,n}\rangle$ in the continuous spectrum is mapped into a solution $|\mathbb{P}_n\rangle$ of the complete (i.e., perturbed) problem, which also belongs to the continuous spectrum and has the same energy $\mathbb{E}_{0,n} = \mathbb{E}_n$. Therefore, implicitly one assumes that the total energy remains constant during the entire process. So, in this manner,

$$\mathbb{H}|\mathbb{P}_n\rangle = (\mathbb{H}_0 + \mathbb{V})|\mathbb{P}_n\rangle = \mathbb{E}_n|\mathbb{P}_n\rangle \quad . \tag{7}$$

We can express the relationship between $|\mathbb{X}_{0,n}\rangle$ and $|\mathbb{P}_n\rangle$ in a more explicit manner by rewriting the previous equations in the form

$$|\mathbb{P}_n^\pm\rangle = |\mathbb{X}_{0,n}\rangle + \mathbb{G}_0^\pm \mathbb{V}|\mathbb{P}_n^\pm\rangle = |\mathbb{X}_{0,n}\rangle + \mathbb{G}^\pm \mathbb{V}|\mathbb{X}_{0,n}\rangle = \hat{\Omega}^\pm|\mathbb{X}_{0,n}\rangle \quad , \tag{8}$$

which is the quaternionic version of the Lippmann-Schwinger equation. Here we have introduced the Moller hyper-operator[20] $\hat{\Omega}^\pm = \mathbb{I} + \mathbb{G}^\pm \mathbb{V} = \mathbb{I} + \mathbb{G}_0^\pm \mathbb{T}^\pm$, with the transition operator $\mathbb{T}^\pm$ being defined as $\mathbb{T}^\pm = \mathbb{V} + \mathbb{V}\mathbb{G}^\pm\mathbb{V}$, where $\mathbb{G}^\pm = \mathbb{G}_0^\pm + \mathbb{G}_0^\pm \mathbb{V}\mathbb{G}^\pm$ is the total Green´s function. Note that $\mathbb{G}_0^\pm$ can be written as

$$\begin{pmatrix} \hat{G}_{01}^\pm & 0 & 0 \\ 0 & \hat{G}_{02}^\pm & 0 \\ 0 & 0 & \hat{G}_{03}^\pm \end{pmatrix} \begin{pmatrix} E - \hat{H}_{01} \pm i\varepsilon & 0 & 0 \\ 0 & E - \hat{H}_{02} \pm j\varepsilon & 0 \\ 0 & 0 & E - \hat{H}_{03} \pm k\varepsilon \end{pmatrix} = \begin{pmatrix} I & 0 & 0 \\ 0 & I & 0 \\ 0 & 0 & I \end{pmatrix} \quad , \tag{9.1}$$

or, in a more compact manner,

$$\mathbb{G}_0^{\pm}(\mathbb{E})\left(\mathbb{E} - \mathbb{H}_0 \pm \mathfrak{q}\,\varepsilon\right) = \mathbb{I}_3 \quad . \tag{9.2}$$

Thus, exactly as in usual Quantum Mechanics[20], $\hat{\Omega}^{\pm}$ transforms each unperturbed basis vector into the corresponding perturbed vector. With the above definition, we are now ready to introduce the scattering matrix $\mathbb{S}$ for quaternionic systems. A scattering event can be described as a transformation of an initial unperturbed immerging state $\left|\mathbb{X}_{0,i}\right\rangle$ into a final (emerging) state $\left|\mathbb{X}_{0,f}\right\rangle$, by effect of a scattering operator $\mathbb{S}$ in the form $\left|\mathbb{X}_{0,f}\right\rangle = \mathbb{S}\left|\mathbb{X}_{0,i}\right\rangle$. In this manner, the amplitude of a given state $\left|\mathbb{X}_{0,n}\right\rangle$ will be given by $\left\langle\mathbb{X}_{0,n}|\mathbb{X}_{0,f}\right\rangle = \left\langle\mathbb{X}_{0,n}|\mathbb{S}|\mathbb{X}_{0,i}\right\rangle = \mathbb{S}_{ni}$. If one takes advantage of the LS equation, one can write $\left|\mathbb{P}_n^{\pm}\right\rangle = \hat{\Omega}^{\pm}\left|\mathbb{X}_{0,n}\right\rangle$, so that the $\mathbb{S}$ matrix can be written as $\mathbb{S} = \hat{\Omega}^{-\dagger}\hat{\Omega}^{+}$, since $\left\langle\mathbb{P}_n^{-}|\mathbb{P}_i^{+}\right\rangle = \left\langle\mathbb{X}_{0,n}|\hat{\Omega}^{-\dagger}\hat{\Omega}^{+}|\mathbb{X}_{0,i}\right\rangle = \left\langle\mathbb{X}_{0,n}|\mathbb{S}|\mathbb{X}_{0,i}\right\rangle$. Thus, the transition probability between two (quaternionic) states will be given by the square of the modulus of the corresponding $\mathbb{S}$ matrix element, i.e., $\wp_{ni} = \left|\mathbb{S}_{ni}\right|^2$.

This general approach allows the application of the quaternionic formalism to different physical situations, such as charge transport in nanoscopic devices and electron transfer in complex systems. For a system that upon a perturbation can evolve alternatively through up to three distinct and non-commutative Hamiltonians, the quaternion formalism permits the knowledge of the total Green´s function, from which one can derive any physical quantity of interest.

To see this, let`s write the total Green´s function $\mathbb{G}^{\pm}$ in the matrix form

$$\mathbb{G}^{\pm} = \begin{pmatrix} \hat{G}_{11}^{\pm} & \hat{G}_{12}^{\pm} & \hat{G}_{13}^{\pm} \\ \hat{G}_{21}^{\pm} & \hat{G}_{22}^{\pm} & \hat{G}_{23}^{\pm} \\ \hat{G}_{31}^{\pm} & \hat{G}_{32}^{\pm} & \hat{G}_{33}^{\pm} \end{pmatrix} \quad (10.1)$$

or, using the standard definitions,

$$\mathbb{G}^{\pm} = \begin{pmatrix} \hat{G}_{11}^{\pm} & \hat{G}_{12}^{\pm} & \hat{G}_{13}^{\pm} \\ \hat{G}_{21}^{\pm} & \hat{G}_{22}^{\pm} & \hat{G}_{23}^{\pm} \\ \hat{G}_{31}^{\pm} & \hat{G}_{32}^{\pm} & \hat{G}_{33}^{\pm} \end{pmatrix} \begin{pmatrix} E - \hat{H}_{01} - \hat{V}_{11} \pm i\varepsilon & -\hat{V}_{12} & -\hat{V}_{13} \\ -\hat{V}_{21} & E - \hat{H}_{02} - \hat{V}_{22} \pm j\varepsilon & -\hat{V}_{23} \\ -\hat{V}_{31} & -\hat{V}_{32} & E - \hat{H}_{03} - \hat{V}_{33} \pm k\varepsilon \end{pmatrix} \quad (10.2)$$

$$= \begin{pmatrix} I & 0 & 0 \\ 0 & I & 0 \\ 0 & 0 & I \end{pmatrix} \quad ,$$

or, finally, in a more compact form,

$$\mathbb{G}^{\pm}(\mathbb{E})(\mathbb{E} - \mathbb{H} \pm \mathfrak{q}\varepsilon) = \mathbb{G}^{\pm}(\mathbb{E})(E\mathbb{I}_3 - \mathbb{H} \pm \mathfrak{q}\varepsilon) = \mathbb{I}_3 \quad . \quad (10.3)$$

Let´s define the Green´s function of the uncoupled perturbed subspaces $\hat{g}_x^{\pm}$ as

$\hat{g}_x^{\pm}(E)(E - \hat{H}_x \pm q_x\varepsilon) = I$ , with $x$ = 1, 2 or 3 (and $q_1$, $q_2$, $q_3 \to i, j, k$). In a matrix form, this yields

$$\begin{pmatrix} \hat{g}_1^{\pm} & 0 & 0 \\ 0 & \hat{g}_2^{\pm} & 0 \\ 0 & 0 & \hat{g}_3^{\pm} \end{pmatrix} \begin{pmatrix} E - \hat{H}_{01} - \hat{V}_{11} \pm i\varepsilon & 0 & 0 \\ 0 & E - \hat{H}_{02} - \hat{V}_{22} \pm j\varepsilon & 0 \\ 0 & 0 & E - \hat{H}_{03} - \hat{V}_{33} \pm k\varepsilon \end{pmatrix} = \quad (11)$$

$$= \begin{pmatrix} I & 0 & 0 \\ 0 & I & 0 \\ 0 & 0 & I \end{pmatrix} \quad .$$

For convenience, from now on we will drop the symbols indicating the advanced and retarded Green´s function. Then, using the above definitions, we can construct the system of equations

$$+\hat{G}_{11}(E - \hat{H}_{01} - \hat{V}_{11}) - \hat{G}_{12}\hat{V}_{21} - \hat{G}_{13}\hat{V}_{31} = I \quad (12.1)$$

$$-\hat{G}_{11}\hat{V}_{12} + \hat{G}_{12}(E - \hat{H}_{02} - \hat{V}_{22}) - \hat{G}_{13}\hat{V}_{32} = 0 \quad (12.2)$$

$$-G_{11}V_{13} - G_{12}V_{23} + G_{13}(E - H_{03} - V_{33}) = 0 \qquad (12.3)$$

that allows to determine all elements of $\mathbb{G}(\mathbb{E})$. For instance, from (12.1), we have $\hat{G}_{11} = \hat{g}_1 + \hat{G}_{12}\hat{V}_{21}\hat{g}_1 + \hat{G}_{13}\hat{V}_{31}\hat{g}_1$, which can be substituted in (12.2), so that $\hat{G}_{12} = \left[\hat{g}_1\hat{V}_{12}\hat{g}_2 + \hat{G}_{13}(\hat{V}_{31}\hat{g}_1\hat{V}_{12}\hat{g}_2 + \hat{V}_{32}\hat{g}_2)\right](I - \hat{V}_{21}\hat{g}_1\hat{V}_{12}\hat{g}_2)^{-1}$.

Now, one can obtain $G_{13}$ by direct substitution,

$$\hat{G}_{13}\left[\left(I - \hat{V}_{31}\hat{g}_1\hat{V}_{13}\hat{g}_3\right) + (\hat{V}_{31}\hat{g}_1\hat{V}_{12}\hat{g}_2 + \hat{V}_{32}\hat{g}_2)(I - \hat{V}_{21}\hat{g}_1\hat{V}_{12}\hat{g}_2)^{-1}\right] = \qquad (13.1)$$
$$= \hat{g}_1\hat{V}_{13}\hat{g}_3 + \hat{g}_1\hat{V}_{12}\hat{g}_2(I - \hat{V}_{21}\hat{g}_1\hat{V}_{12}\hat{g}_2)^{-1}\left(\hat{V}_{21}\hat{g}_1\hat{V}_{13}\hat{g}_3 + \hat{V}_{23}\hat{g}_3\right)$$

$$\hat{G}_{13} = \left[\hat{g}_1\hat{V}_{13}\hat{g}_3 + \hat{g}_1\hat{V}_{12}\hat{g}_2(I - \hat{V}_{21}\hat{g}_1\hat{V}_{12}\hat{g}_2)^{-1}\left(\hat{V}_{21}\hat{g}_1\hat{V}_{13}\hat{g}_3 + \hat{V}_{23}\hat{g}_3\right)\right] \times$$
$$\times \left[\left(I - \hat{V}_{31}\hat{g}_1\hat{V}_{13}\hat{g}_3\right) + (\hat{V}_{31}\hat{g}_1\hat{V}_{12}\hat{g}_2 + \hat{V}_{32}\hat{g}_2)(I - \hat{V}_{21}\hat{g}_1\hat{V}_{12}\hat{g}_2)^{-1}\left(\hat{V}_{21}\hat{g}_1\hat{V}_{13}\hat{g}_3 + \hat{V}_{23}\hat{g}_3\right)\right]^{-1} \quad (13.2)$$

Once $\hat{G}_{13}$ and $\hat{G}_{12}$ are known, $\hat{G}_{11}$ can be also determined, and, in fact, one can apply the same reasoning to calculate all other terms that appear in $\mathbb{G}(\mathbb{E})$. No approximations are involved, and note that all expressions depend only of the exact Green´s function of the uncoupled quaternionic sub-spaces ($\hat{g}_x$), which admits the usual expansion in a Dyson series of the type $\hat{g}_x = \hat{G}_{0x} + \hat{G}_{0x}\hat{V}_{xx}\hat{g}_x$.

However, by retaining terms up to only a certain order in $\hat{g}_x$, one can obtain simpler expressions corresponding to non-exact expressions for $\mathbb{G}(\mathbb{E})$ that are useful for practical reasons. For example, in a treatment similar to that adopted in the usual first Born approximation[4] for $\hat{g}_x$, we can introduce a "hyper" first Born approximation for $\mathbb{G}(\mathbb{E})$ by preserving only terms that involve the exact Green´s function of the uncoupled quaternionic sub-spaces

$$\mathbb{G}(\mathbb{E}) \approx \begin{pmatrix} \hat{g}_1 & 0 & 0 \\ 0 & \hat{g}_2 & 0 \\ 0 & 0 & \hat{g}_3 \end{pmatrix} \quad . \tag{14}$$

We will now show that in this case the relationship between $\mathbb{S}$ and $\mathbb{T}$ is similar to that found in usual Quantum Mechanics.

### B. Relationship between the scattering hypermatrix and the transition matrix

A formal relationship between the $\mathbb{S}$ and $\mathbb{T}$ matrices can be directly obtained in the form

$$\begin{aligned}
\mathbb{S}_{fi} &= \langle \mathbb{P}_f^- | \mathbb{P}_i^+ \rangle = \langle \mathbb{P}_f^- | \hat{\Omega}^+(E_i) | \mathbb{X}_{0,i} \rangle \\
&= \langle \mathbb{P}_f^- | \mathbb{I} + \mathbb{G}^+(E_i) \mathbb{V} | \mathbb{X}_{0,i} \rangle \\
&= \langle \mathbb{P}_f^- | \mathbb{X}_{0,i} \rangle + \langle \mathbb{P}_f^- | \mathbb{G}^+(E_i) \mathbb{V} | \mathbb{X}_{0,i} \rangle \\
&= \langle \mathbb{X}_{0,f} | \mathbb{X}_{0,i} \rangle + \langle \mathbb{X}_{0,f} | \mathbb{T}^{-\dagger}(E_f) \mathbb{G}_0^{-\dagger}(E_f) | \mathbb{X}_{0,i} \rangle + \langle \mathbb{P}_f^- | \mathbb{G}^+(E_i) \mathbb{V} | \mathbb{X}_{0,i} \rangle \quad .
\end{aligned} \tag{15}$$

Now, if we use the identities $\mathbb{I} = |\tilde{\mathbb{X}}_{0,s}\rangle\langle\tilde{\mathbb{X}}_{0,s}| = \begin{pmatrix} \mathbb{I}_1 & 0 & 0 \\ 0 & \mathbb{I}_2 & 0 \\ 0 & 0 & \mathbb{I}_3 \end{pmatrix} \neq |\mathbb{X}_{0,s}\rangle\langle\mathbb{X}_{0,s}|$ and $\mathbb{I} = |\tilde{\mathbb{P}}_f^-\rangle\langle\tilde{\mathbb{P}}_f^-|$, we have

$$\begin{aligned}
\mathbb{S}_{fi} &= \hat{\delta}_{fi} + \langle \mathbb{X}_{0,f} | \mathbb{T}^{-\dagger}(E_f) | \tilde{\mathbb{X}}_{0,\bar{i}} \rangle\langle \tilde{\mathbb{X}}_{0,\bar{i}} | \mathbb{G}_0^{-\dagger}(E_f) | \mathbb{X}_{0,i} \rangle + \langle \mathbb{P}_f^- | \mathbb{G}^+(E_i) | \tilde{\mathbb{P}}_{\bar{f}}^- \rangle\langle \tilde{\mathbb{P}}_{\bar{f}}^- | \mathbb{V} | \mathbb{X}_{0,i} \rangle \\
&= \hat{\delta}_{fi} + \langle \mathbb{X}_{0,f} | \mathbb{T}^{-\dagger}(E_f) | \tilde{\mathbb{X}}_{0,\bar{i}} \rangle\langle \tilde{\mathbb{X}}_{0,\bar{i}} | \mathbb{G}_0^{-\dagger}(E_f) | \mathbb{X}_{0,i} \rangle + \langle \mathbb{P}_f^- | \mathbb{G}^+(E_i) | \tilde{\mathbb{P}}_{\bar{f}}^- \rangle\langle \tilde{\mathbb{X}}_{0,\bar{f}} | \hat{\Omega}^{-\dagger}(E_f) \mathbb{V} | \mathbb{X}_{0,i} \rangle \\
&= \hat{\delta}_{fi} + \langle \mathbb{X}_{0,f} | \mathbb{T}^{-\dagger}(E_f) | \tilde{\mathbb{X}}_{0,\bar{i}} \rangle\langle \tilde{\mathbb{X}}_{0,\bar{i}} | \mathbb{G}_0^{-\dagger}(E_f) | \mathbb{X}_{0,i} \rangle + \langle \mathbb{P}_f^- | \mathbb{G}^+(E_i) | \tilde{\mathbb{P}}_{\bar{f}}^- \rangle\langle \tilde{\mathbb{X}}_{0,\bar{f}} | \mathbb{T}^{-\dagger}(E_f) | \mathbb{X}_{0,i} \rangle \quad .
\end{aligned} \tag{16}$$

In the above expression, we note that while the matrix $\mathbb{G}_0^{-\dagger}$ is diagonal, $\mathbb{G}^+$ has a more complicated structure. However, within the $\mathbb{G}^+(\mathbb{E})$ approximation (see Eqs. (13) and (14)) the relationship between $\mathbb{S}$ and $\mathbb{T}$ assumes the form

$$\mathbb{S}_{fi} \cong \hat{\delta}_{fi} + \langle \mathbb{X}_{0,f} | \mathbb{T}^{-\dagger}(E_f) | \tilde{\mathbb{X}}_{0,\bar{i}} \rangle \langle \tilde{\mathbb{X}}_{0,\bar{i}} | \mathbb{G}_0^{-\dagger}(E_f) | \mathbb{X}_{0,i} \rangle + \qquad (17.1)$$
$$+ \langle \tilde{\mathbb{P}}_f^- | \mathbb{G}^+(E_i) | \tilde{\mathbb{P}}_{\bar{f}}^- \rangle \langle \tilde{\mathbb{X}}_{0,\bar{f}} | \mathbb{T}^{-\dagger}(E_f) | \mathbb{X}_{0,i} \rangle$$

$$\cong \hat{\delta}_{fi} + \langle \mathbb{X}_{0,f} | \mathbb{T}^{-\dagger}(E_f) | \tilde{\mathbb{X}}_{0,\bar{i}} \rangle \langle \tilde{\mathbb{X}}_{0,\bar{i}} | \mathbb{G}_0^{-\dagger}(E_f) \Upsilon | \tilde{\mathbb{X}}_{0,i} \rangle + \qquad (17.2)$$
$$+ \langle \tilde{\mathbb{P}}_f^- | \Upsilon \mathbb{G}^+(E_i) | \tilde{\mathbb{P}}_{\bar{f}}^- \rangle \langle \tilde{\mathbb{X}}_{0,\bar{f}} | \mathbb{T}^{-\dagger}(E_f) | \mathbb{X}_{0,i} \rangle$$

$$\cong \hat{\delta}_{fi} + \langle \mathbb{X}_{0,f} | \mathbb{T}^{-\dagger}(E_f) \Upsilon | \tilde{\mathbb{X}}_{0,\bar{i}} \rangle \langle \tilde{\mathbb{X}}_{0,\bar{i}} | \mathbb{G}_0^{-\dagger}(E_f) | \tilde{\mathbb{X}}_{0,i} \rangle + \qquad (17.3)$$
$$+ \langle \tilde{\mathbb{P}}_f^- | \mathbb{G}^+(E_i) | \tilde{\mathbb{P}}_{\bar{f}}^- \rangle \langle \tilde{\mathbb{X}}_{0,\bar{f}} | \Upsilon \mathbb{T}^{-\dagger}(E_f) | \mathbb{X}_{0,i} \rangle$$

$$\cong \hat{\delta}_{fi} + \langle \mathbb{X}_{0,f} | \mathbb{T}^{-\dagger}(E_f) | \mathbb{X}_{0,\bar{i}} \rangle \hat{\delta}_{\bar{i}i} \left( \mathbb{G}_0^{-\dagger}(E_f) \right)_{\bar{i}i} + \qquad (17.4)$$
$$+ \langle \mathbb{X}_{0,\bar{f}} | \mathbb{T}^{-\dagger}(E_f) | \mathbb{X}_{0,i} \rangle \hat{\delta}_{f\bar{f}} \left( \mathbb{G}^+(E_i) \right)_{f\bar{f}}$$

$$\cong \hat{\delta}_{fi} + \langle \mathbb{X}_{0,f} | \mathbb{T}^{-\dagger}(E_f) | \mathbb{X}_{0,i} \rangle \left[ \left( \mathbb{G}_0^{-\dagger}(E_f) \right)_{ii} + \left( \mathbb{G}^+(E_i) \right)_{ff} \right] \qquad (17.5)$$

where we have used

$$|\mathbb{X}_{0,n}(t)\rangle \equiv \begin{pmatrix} \gamma_1 |\phi_{n_1}(t)\rangle \\ \gamma_2 |\phi_{n_2}(t)\rangle \\ \gamma_3 |\phi_{n_3}(t)\rangle \end{pmatrix} = \begin{pmatrix} \gamma_1 & 0 & 0 \\ 0 & \gamma_2 & 0 \\ 0 & 0 & \gamma_3 \end{pmatrix} \begin{pmatrix} |\phi_{n_1}(t)\rangle \\ |\phi_{n_2}(t)\rangle \\ |\phi_{n_3}(t)\rangle \end{pmatrix} = \Upsilon |\tilde{\mathbb{X}}_{0,n}(t)\rangle \quad . \qquad (18)$$

If we consider the term inside the brackets in an explicit manner,

$$\left[ \left( \mathbb{G}_0^{-\dagger}(E_f) \right)_{ii} + \left( \mathbb{G}^+(E_i) \right)_{ff} \right] =$$

$$= Tr \left[ \begin{pmatrix} \frac{1}{E_{f1}-E_{i1}+i\eta} & 0 & 0 \\ 0 & \frac{1}{E_{f2}-E_{i2}+j\eta} & 0 \\ 0 & 0 & \frac{1}{E_{f3}-E_{i3}+k\eta} \end{pmatrix} - \begin{pmatrix} \frac{1}{E_{f1}-E_{i1}-i\eta} & 0 & 0 \\ 0 & \frac{1}{E_{f2}-E_{i2}-j\eta} & 0 \\ 0 & 0 & \frac{1}{E_{f3}-E_{i3}-k\eta} \end{pmatrix} \right] \qquad (19)$$

$$= -2\pi \left[ i\delta(E_{f1} - E_{i1}) + j\delta(E_{f2} - E_{i2}) + k\delta(E_{f3} - E_{i3}) \right] \quad ,$$

we obtain

$$\mathbb{S}_{fi} \cong \hat{\delta}_{fi} - 2\pi \left[ i\delta(E_{f1} - E_{i1}) + j\delta(E_{f2} - E_{i2}) + k\delta(E_{f3} - E_{i3}) \right] \left( \mathbb{T}^{-\dagger}(E_f) \right)_{fi}. \qquad (20)$$

It is reassuring to note that if we take the limit where $\gamma_1 = 1$ and $\gamma_2 = \gamma_3 = 0$ in the above expression, we obtain the correct result for the case where a single subspace is considered, i.e., $S_{fi} = \delta_{fi} - 2\pi i \delta(E_f - E_i)\left(T^{-\dagger}(E_f)\right)_{fi}$. Also, when $f \neq i$ and the initial and final states belong to the same continuous spectrum of energy, we have $E_{f1} = E_{f2} = E_{f3} = E_f$ and $E_{i1} = E_{i2} = E_{i3} = E_i$. In this manner, finally we have $\mathbb{S}_{fi} \cong -2\pi\delta(E_f - E_i)[i + j + k]\left(\mathbb{T}^{-\dagger}(E_f)\right)_{fi}$, whose modulus squared corresponds to the transition probability $\wp_{fi} = |\mathbb{S}_{fi}|^2$, i.e.,

$$\left|\mathbb{S}_{fi}\right|^2 = 12\pi^2 \delta^2(E_f - E_i)\left|\left(\mathbb{T}^{\dagger}(E_f)\right)_{fi}\right|^2 = 12\pi^2 \delta(0)\delta(E_f - E_i)\left|\left(\mathbb{T}(E_f)\right)_{fi}\right|^2. \tag{21}$$

However, since $\delta(0) = \lim_{T \to \infty}\left(\dfrac{T}{2\pi\hbar}\right)$, we can write the transition rate between the final and initial states ($\mathbb{R}_{fi}$) as

$$\frac{d\wp_{fi}}{dT} = \mathbb{R}_{fi} = 6\pi\hbar^{-1}\delta(E_f - E_i)\left|\left(\mathbb{T}(E_f)\right)_{fi}\right|^2. \tag{22}$$

We can easily recognize the above expression as Fermi golden rule generalized for quaternionic systems.

## IV. AN EXPRESSION FOR THE ELECTRIC CURRENT

The generalized Fermi Golden Rule can be used to estimate the electric current and, consequently, the conductance of nanoscopic systems. For this, let us consider that the continuous spectrum of initial [final] states as associated to the right ($i \to R$) [left ($f \to L$)] electrode. We also have that each quaternionic subspace represents a possible charge state (i.e., neutral, single cation, single anion) of the device. If we multiply the above expression by the charge *e*, and transform the sums over the final and initial states into integrals ( $\sum_{\beta=L(R)} \to \int dE_\beta \, \rho_{(E_\beta)}$, where $\rho_{(E_\beta)}$ is the density of states with $\beta = L, R$), we have

$$I_{RL} = 6e\pi\hbar^{-1} \iint dE_L \, dE_R \left[ \rho_{(E_R)} \rho_{(E_L)} \delta(E_L - E_R) \left| \left( \mathbb{T}_{(E_R)} \right)_{RL} \right|^2 \right] \quad . \tag{23}$$

In the specific case of electronic transport in nanoscopic systems (such as when a single molecule is connected to two electrodes, for instance), it is common to consider the fact that no direct transport exists between the two electrodes, but only through the molecule. Hence, the right to left (R→L) transmission will be given by

$$\begin{aligned} \left| \left( \mathbb{T}_{(E_R)} \right)_{RL} \right|^2 &= \left( \mathbb{T}^+_{(E_R)} \right)_{RL} \left( \mathbb{T}^+_{(E_R)} \right)_{RL}^\dagger \\ &= \left\langle \mathbb{X}_{0,L} \left| \mathbb{V}_L \mathbb{G}^+_{(E_R)} \mathbb{V}_R \right| \mathbb{X}_{0,R} \right\rangle \left\langle \mathbb{X}_{0,R} \left| \mathbb{V}_R^\dagger \mathbb{G}^-_{(E_R)} \mathbb{V}_L^\dagger \right| \mathbb{X}_{0,L} \right\rangle \end{aligned} \quad . \tag{24}$$

If we use again the identity $\mathbb{I}_N = \sum_N \left| \tilde{\mathbb{X}}_{0,N} \right\rangle \left\langle \tilde{\mathbb{X}}_{0,N} \right|$ and take into account the invariance of the trace, we obtain

$$\begin{aligned} I_{RL} &= \frac{6e\pi}{\hbar} \sum_N \iint dE_L \, dE_R \left[ \rho_{(E_R)} \rho_{(E_L)} \delta(E_L - E_R) \left\langle \mathbb{X}_{0,L} \left| \mathbb{V}_L \mathbb{G}^+_{(E_R)} \mathbb{V}_R \right| \mathbb{X}_{0,R} \right\rangle \cdot \right. \\ &\qquad \left. \cdot \left\langle \mathbb{X}_{0,R} \left| \mathbb{V}_R^\dagger \mathbb{G}^-_{(E_R)} \right| \tilde{\mathbb{X}}_{0,N} \right\rangle \left\langle \tilde{\mathbb{X}}_{0,N} \left| \mathbb{V}_L^\dagger \right| \mathbb{X}_{0,L} \right\rangle \right] \\ &= \frac{6e\pi}{\hbar} \sum_N \iint dE_L \, dE_R \left[ \rho_{(E_R)} \rho_{(E_L)} \delta(E_L - E_R) \left\langle \tilde{\mathbb{X}}_{0,N} \left| \mathbb{V}_L^\dagger \right| \mathbb{X}_{0,L} \right\rangle \cdot \right. \\ &\qquad \left. \cdot \left\langle \mathbb{X}_{0,L} \left| \mathbb{V}_L \mathbb{G}^+_{(E_R)} \mathbb{V}_R \right| \mathbb{X}_{0,R} \right\rangle \left\langle \mathbb{X}_{0,R} \left| \mathbb{V}_R^\dagger \mathbb{G}^-_{(E_R)} \right| \tilde{\mathbb{X}}_{0,N} \right\rangle \right] \end{aligned} \tag{25}$$

After integrating in $E_L$, and considering $E_R = E$, we get

$$I_{RL} = \frac{6e}{4\pi\hbar} \sum_N \int dE \left[ \left\langle \tilde{\mathbb{X}}_{0,N} \left| \left( 2\pi\rho_{(E)} \mathbb{V}_L^\dagger \right| \mathbb{X}_{0,L} \right\rangle \left\langle \mathbb{X}_{0,L} \right| \mathbb{V}_L \right) \bullet \right. \\ \left. \bullet \, \mathbb{G}^+(E) \left( 2\pi\rho_{(E)} \mathbb{V}_R \left| \mathbb{X}_{0,R} \right\rangle \left\langle \mathbb{X}_{0,R} \right| \mathbb{V}_R^\dagger \right) \mathbb{G}^-(E) \left| \tilde{\mathbb{X}}_{0,N} \right\rangle \right] \quad . \tag{26}$$

If we define the matrix: $\mathbb{A}_w = \left(2\pi\rho_{(E)} \mathbb{V}_w^\dagger |\mathbb{X}_{0,w}\rangle\langle\mathbb{X}_{0,w}| \mathbb{V}_w\right)$ ($w$ = R, L), sum over the $N$ states and consider that the incident electron goes from an occupied level belonging to the right electrode (and, hence, described by the Fermi occupation factor $f(E - \mu_R)$) to an unoccupied level in the left electrode (described by the Fermi factor $A(E - \mu_L) = 1 - f(E - \mu_L)$), we have

$$I_{RL} = \frac{3e}{h} \int dE\, \hat{T}r\left\{\mathbb{A}_L \mathbb{G}^+(E) \mathbb{A}_R \mathbb{G}^-(E)\right\} f(E - \mu_R) A(E - \mu_L) \quad, \tag{27}$$

where the symbol $\hat{T}r$ denotes the trace over a quaternionic function ($|\tilde{\mathbb{X}}_{0,N}\rangle$). In an analogous manner, the left to right current will be

.

$$I_{LR} = \frac{3e}{h} \int dE\, \hat{T}r\left\{\mathbb{A}_L \mathbb{G}^+(E) \mathbb{A}_R \mathbb{G}^-(E)\right\} A(E - \mu_R) f(E - \mu_L) \quad. \tag{28}$$

Hence, the total current can be written as

$$I_T = I_{RL} - I_{LR} = \frac{3e}{h} \int dE\, \hat{T}r\left\{\mathbb{A}_L \mathbb{G}^+(E) \mathbb{A}_R \mathbb{G}^-(E)\right\} \left(f(E - \mu_R) - f(E - \mu_L)\right) \quad, \tag{29}$$

which is the quaternionic version of the Landauer formula. In addition, we can identify the transmission as

$$\Im(E) = \hat{T}r\left\{\mathbb{A}_L \mathbb{G}^+(E) \mathbb{A}_R \mathbb{G}^-(E)\right\} \quad. \tag{30}$$

Although the above expression is quite similar to that used to describe the ballistic transport (see Eq. (1)), the present formalism makes possible to consider three alternative charge states for the system, in such a way that the evolution of the transport process can naturally lead to the mixing of the three different quaternionic subspaces. This intrinsic feature of the treatment emerges because the off diagonal elements of the matrix quantity $\mathbb{A}_w = \left(2\pi\rho_{(E)} \mathbb{V}_w^\dagger |\mathbb{X}_{0,w}\rangle\langle\mathbb{X}_{0,w}| \mathbb{V}_w\right)$ (where $w$ = R, L) introduce the mixing

of the subspaces. (Please note that this occurs in spite of the approximation done for $\mathbb{G}(\mathbb{E})$ in Eq. (14), when only the (exact) diagonal terms of the total Green´s function of each subspace were considered.) Thus, for instance, one electron could enter a system that was in an initially neutral state, but leave it when the system is in a negatively charged state. Because the total energy is conserved, the two states involved (one at the neutral and the other at the charged species) must be identical in energy and readily accessible; at least, if and only if broadening effects are included, nearby energy states should be available to participate of the transport process. This hypothetical, but possible, situation is depicted in Figure 1.

## V. APPLICATIONS: TRANSPORT IN CONJUGATED AND SATURATED SYSTEMS

As a stringent test of the range of applications of the theoretical formalism developed above, we examined two archetypical molecules of different kinds: benzene-di-thiol (BDT), a conjugated system, and octane-di-thiol (ODT), whose structure only contains saturated bonds. We have chosen these molecules not only because they represent canonical examples of strong and weak coupling regimes, respectively,[10, 21, 22] but also because experimental data for the current and conductance exist for them.[23-25] Here, we will briefly present the most relevant results (see Figs. 2-4), since a more detailed discussion can be found in Ref. [26].

In Figs. 2 and 3 we not only depict the corresponding extended molecules, where two opposite ends of the organic compound considered was coupled to a metallic cluster of 11 gold atoms, but also show some selected frontier molecular orbitals at some special values of the external potential that most affect them. This extended molecule was used to extract the coupling strengths at DFT level of calculation using the B3LYP functional and the 6-31d basis set, for different values of an external electric field. The curves representing the behavior of the current and the conductance of these two molecules are shown in Figs. 4 and 5, where in each case we also indicate the individual contributions of the ballistic and co-tunneling terms to the total current traversing the molecule.

As one can see in Fig. 2b, in the case of the BDT molecule, up to an applied voltage $V_{appl}$ = 0.84 V the LUMO is completely localized in the left-hand side cluster. However, at this voltage an avoided-crossing situation[27] begins to occur for the HOMO and LUMO (as depicted in the inset of Fig. 4b), resulting in an exchange of the spatial localization of these two FMOs. Above this critical limit of the applied potential, the LUMO begins to act as a true "conducting channel" for the system, a fact that is associated to the now increasing values of the up to then non-existent electric current. The most noticeable increase of the current occurs at an external bias of 1.1V, when the Fermi level crosses the energy of the LUMO+1, a completely delocalized orbital.

As expected, an entirely different physical situation is responsible for the transport in the saturated molecule ODT, for which no avoided-crossing situation (and, therefore, no consequent symmetry exchange) should be anticipated to occur. Since the LUMO remains localized in the left-hand side cluster for the entire range of examined potentials, it cannot contribute to a ballistic transport. This spatial localization precludes the transfer of electrons from the metal to the molecule. Hence, with no coupling between the cluster and the molecule itself, it is unlikely that an anion could be formed. However, as one can see in Fig. 3b, the HOMO – which is localized both in the organic molecule as in the right-hand side cluster – can lose one electron to the anode. As the cation is formed, the local density of states is changed and now the unoccupied molecular orbitals of the newly formed electron-deficient molecular system can accept electrons from the cathode. Hence, a typical co-tunneling situation develops accordingly to what can be seen in Fig. 5b, where the contribution of the ballistic terms does not exist for the entire range of applied external field.

In general, the above results are in a good qualitative agreement with the experimental data available for these systems.[24, 25] The interested reader can find a more complete discussion of the application of the quaternionic formalism to the description of the transport in these two model systems in Ref. [26].

## VI.  CONCLUSIONS AND PERSPECTIVES

We have presented the quaternionic method, a new theoretical approach specially developed to deal with the calculation of the electric current that flows through a molecule subject to an externally applied electric field, while connected to two terminal electrodes. A full ab initio self-consistent calculation can be implemented for the so-called extended molecule, which comprises the organic molecule of interest plus the two small metal clusters attached to it. The quarternionic approach allows for the possibility of concomitant ballistic and co-tunneling transport mechanisms, and the molecular conductance profile is particularly sensitive to the opening and closing of transport channels through the transient single-charged species. For this, we have chosen to follow a Dyson series and a generalized Fermi golden rule treatment within a quaternion formalism, where the evolution of an initially neutral molecule can proceed either preserving the neutral charge or by following the Hamiltonians that describe a single-charged anion or a single-charged cation. In this manner, one can go beyond a canonical quantum mechanical formalism of the separated subsystems, since we allow for the possibility of coupling between the three quaternionic subspaces (each one with its own imaginary unit and corresponding to the three different charge states of the molecular system).

The main advantage of the present approach is that in the final expression for the transmission of an electron flowing through the molecule, the ballistic and co-tunneling terms appear naturally as complementary regimes of transport. Also, the treatment can take into account the possibility of describing strongly correlated situations (as those associated to the presence of spatially localized molecular orbitals) where the ballistic regime usually fails.[3] The allowance for a natural "competition" between the ballistic and co-tunneling regimes gives rise to the possibility of application of the quaternion-based formalism to strong correlated systems where atomic-like localized molecular states play an important role.

As illustrative examples of the range of applications of the quaternionic approach, we have briefly discussed the transport in the organic molecules BDT and ODT, whose field-dependent electronic structures were determined using the B3LYP functional approximation and adopting the 6-31d basis set. For the first molecule, a conjugated system, we have found the ballistic mechanism to be dominant, as it should be expected. For the second, whose unoccupied states are spatially localized, the transport is made possible by the transfer of one electron from the molecule to the anode, with the frontier molecular orbitals of the corresponding cation playing an active role in the process.

We have shown that the present formalism allows for a unified description of the transport phenomena in molecular systems. Besides the ballistic processes, a non-zero probability exists for the capture of an electron or a hole from the connected electrodes, so that a change the charge of the molecule would allow the charge transfer to proceed in a new potential surface. The coexistence of different mechanisms not only gives opportunity to new possibilities of approaching and interpreting the molecular conductance problem, but also opens up several evident possibilities of extension of the present work.

From the point of view of the quaternion formalism, one can imagine its application to the description of other physical situations where more than one evolution pathway is possible, as it would be the case of describing co-existing molecular tautomers or isomers. Furthermore, in case where more than three viable pathways for the time evolution of a given system can be foreseen, one could consider resorting to octanions, as to accommodate a larger gamut of different possibilities.

References

1. A. Aviram and M. A. Ratner, Chemical Physics Letters 29 (2), 277-283 (1974).
2. C. Durkan, Current at the nanoscale: an introduction to nanoelectronics. (Imperial College Press; Distributed by World Scientific Pub., London, 2007).
3. S. Datta, Electronic transport in mesoscopic systems. (Cambridge University Press, Cambridge; New York, 1995).
4. M. Di Ventra, Electrical transport in nanoscale systems. (Cambridge University Press, Cambridge, 2008).
5. E. S. C. C. Juan, Molecular Electronics: An Introduction to Theory and Experiment (Word Scientific, 2010).
6. S. Datta, Quantum transport: atom to transistor. (Cambridge University Press, Cambridge, UK ; New York, 2005).
7. C. K. Wang, Y. Fu and Y. Luo, Physical Chemistry Chemical Physics 3 (22), 5017-5023 (2001).
8. M. Lundstrom, Nanoscale transistors: device physics, modeling and simulation. (Springer, New York, 2006).
9. S. Datta, M. A. Ratner and Y. Xue, Chemical Physics 281, 20 (2002).
10. K. Stokbro, J. Taylor, M. Brandbyge and H. Guo, in Introducing Molecular Electronics, edited by T. L. N. O. Physics (Springer, Berlin, 2005), Vol. 680, pp. 34.
11. M. A. Ratner, A. Nitzan and M. Galperin, Abstracts of Papers of the American Chemical Society 230, U2775-U2775 (2005).
12. D. K. Ferry and S. M. Goodnick, Transport in nanostructures. (Cambridge University Press, Cambridge, U.K.; New York, 1997).
13. R. A. Moreira and C. P. de Melo, submitted to The Journal of Chemical Physics (2014).
14. M. D. Ventra, Electrical Transport in Nanoscale Systems. (Cambridge University Press, New York, 2008).
15. F. Remacle and R. D. Levine, Chemical Physics Letters 383 (5-6), 537-543 (2004).
16. R. Li, S. M. Hou, J. X. Zhang, Z. K. Qian, Z. Y. Shen and X. Y. Zhao, J. Chem. Phys. 125 (19) (2006).
17. D. V. Averin and Y. V. Nazarov, Physical Review Letters 65 (19), 4 (1990).
18. W. R. Hamilton, Elements of Quaternions. (Dublin, 1866).
19. S. L. Adler, Quaternionic Quantum Mechanics and Quantum Fields. (Oxford University Press, 1995).
20. R. D. Levine, Quantum Mechanics of Molecular Rate Process. (Dover, New York, 1969).
21. M. Lundstrom and J. Guo, Nanoscale transistors: device physics, modeling and simulation. (Springer, New York, 2006).
22. F. Jiang, Y. X. Zhou, H. Chen, R. Note, H. Mizuseki and Y. Kawazoe, Physical Review B 72 (15), 155408 (2005).
23. C.-L. Ma, D. Nghiem and Y.-C. Chen, Applied Physics Letters 93 (22), 222111-222113 (2008).
24. C. A. Martin, D. Ding, H. S. J. van der Zant and J. M. van Ruitenbeek, New Journal of Physics 10 (6), 065008 (2008).
25. J. Tomfohr, G. K. Ramachandran, O. F. Sankey and S. M. Lindsay, in Introducing Molecular Electronics, edited by G. Cuniberti, K. Richter and G. Fagas (Springer Berlin / Heidelberg, 2005), Vol. 680, pp. 301-312.

26. A. C. L. Moreira and C. P. de Melo, to be submitted (2014).
27. C. P. de Melo and A. C. L. Moreira, The Journal of Physical Chemistry C 116 (4), 3122-3131 (2011).

Figure Captions

Figure 1: Two possible situations where only the subspaces 1 (left panel) and 2 (right panel) are considered. Each subspace (1 and 2) is composed by a two-level system (levels *a* and *b*). The symbol $\mu_\sigma$ represents the chemical potential of the $\sigma$ - electrode ($\sigma = L, R$), $\varepsilon_{xy}$ the *x*-level of the *y*-subspace (i.e., *x* = *a*, *b* and *y* = 1, 2), and $\tau_{xy}^\sigma$ is the coupling between this energy level and the $\sigma$ - electrode. Due to the coupling to the infinite electrodes, all levels (*a* and *b*) in all subspaces (1 and 2) are broadened.

Figure 2: (a) The extended molecule benzene-di-thiol where each metallic cluster contains 11 gold atoms. (b) Behavior of selected molecular orbitals as a function of the applied voltage.

Figure 3: (a) The extended molecule octane-di-thiol where each metallic cluster contains 11 gold atoms. (b) Behavior of selected molecular orbitals as a function of the applied voltage.

Figure 4: Electric current and conductance (a) and ballistic and co-tunneling contributions to the current (b) for benzene-di-thiol as a function of the applied voltage. Inset: HOMO-LUMO crossing.

Figure 5: Electric current and conductance (a) and ballistic and co-tunneling contributions to the current (b) for octane-di-thiol as a function of the applied voltage.

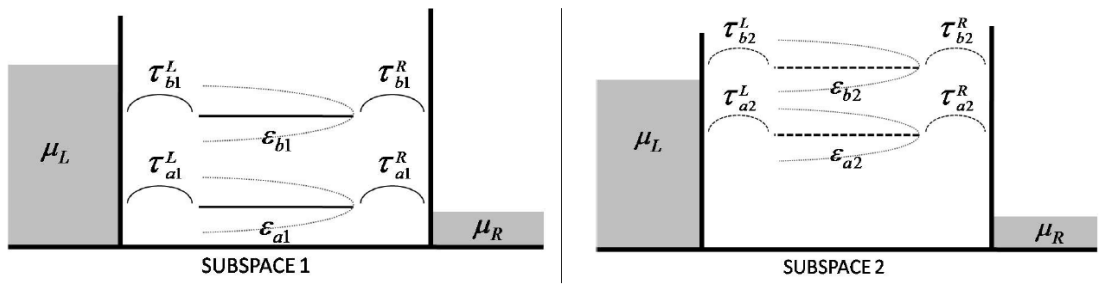

**Figure 1**

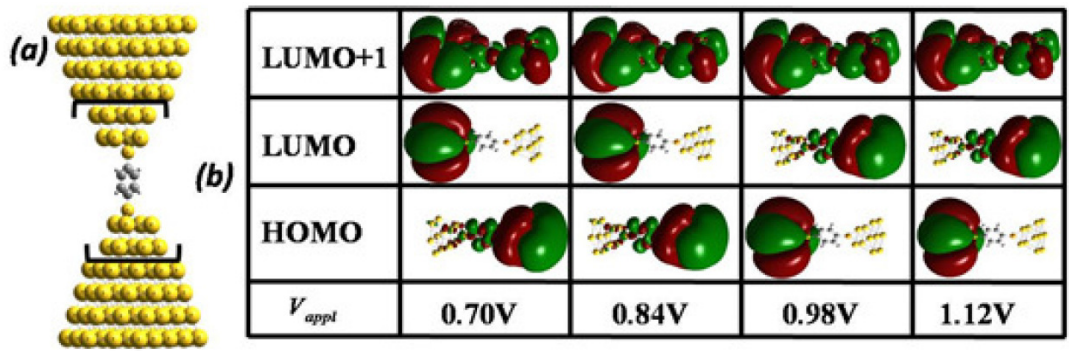

**Figure 2**

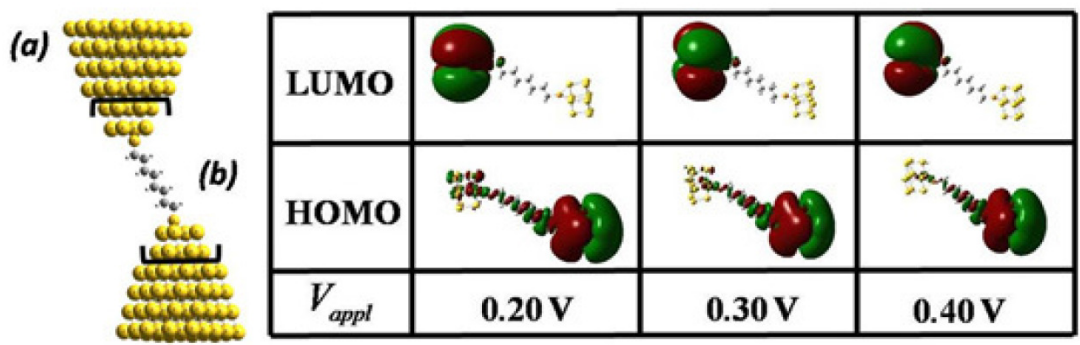

**Figure 3**

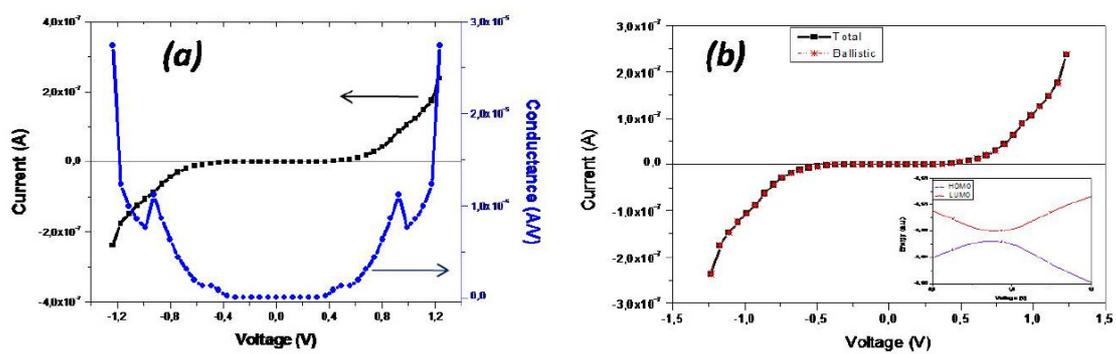

**Figure 4**

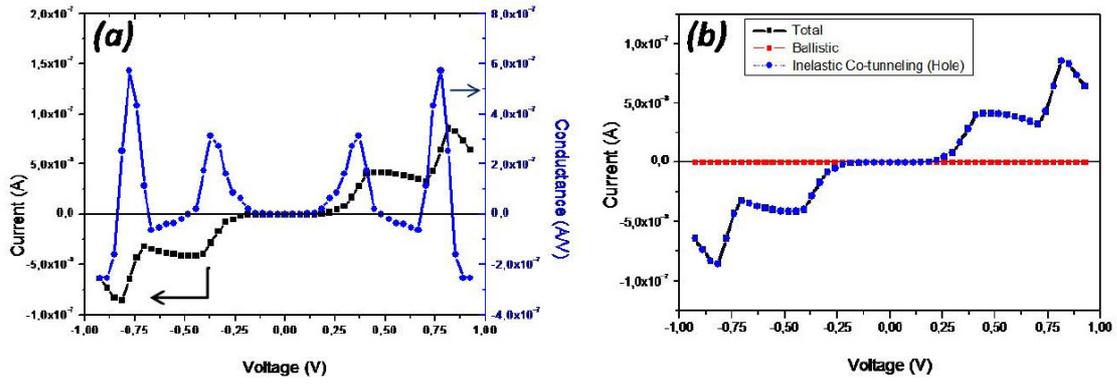

**Figure 5**